\documentclass[preprint,12pt]{elsarticle}

\usepackage{amssymb}
\usepackage{amsmath}

\journal{Computers And Fluids}

\begin{document}

\begin{frontmatter}



\title{Convolutional Neural Networks For Turbulent Model Uncertainty Quantification}

\author[inst1]{Minghan Chu}

\affiliation[inst1]{organization={Mechanics and Materials Engineering Department},
            addressline={130 Stuart Street}, 
            city={Kingston},
            postcode={K7L 3N6}, 
            state={ON},
            country={Canada}}

\author[inst2]{Weicheng Qian}

\affiliation[inst2]{organization={Department of Computer Science, University of Saskatchewan},
            addressline={176 Thorvaldson Bldg, 110 Science Place}, 
            city={Saskatoon},
            postcode={S7N 5C9}, 
            state={SK},
            country={Canada}}

\begin{abstract}
Complex turbulent flow simulations are an integral aspect of the engineering design process. The mainstay of these simulations is represented by eddy viscosity based turbulence models. Eddy viscosity models are computationally cheap due to their underlying simplifications, but their predictions are also subject to structural errors. At the moment, the only method available to forecast these uncertainties is the Eigenspace Perturbation Method. This method's strictly physics-based approach frequently results in unreasonably high uncertainty bounds, which drive the creation of extremely cautious designs. To tackle this problem, we employ a strategy based on deep learning. In order to control the perturbations, our trained deep learning models forecast the appropriate amount of disturbance to apply to the anticipated Reynolds stresses.  A Convolutional Neural Network is used to carry out this, and it is trained to distinguish between high fidelity data, which is a mapping of flow characteristics, and model projections based on eddy viscosity. 
\end{abstract}




\end{frontmatter}


\section{Background And Motivation}
A significant issue in many different engineering design challenges is to account for the effects of fluid turbulence and the evolution of turbulent flows. Despite much investigation, no analytical theory has been able to accurately predict the evolution of complicated turbulent flows found in real life designs. Turbulence models must be used in engineering design research for simulations. These constitutive equations for turbulence models are simplified and connect difficult-to-compute quantities of relevance to known, readily-computable ones. 

For engineering design, a variety of turbulence modeling techniques are available, from Eddy Viscosity Models (EVM) \cite{craft1996development, gatski2002linear, shih1995new, kraichnan1976eddy} to Large Eddy Simulations (LES) \cite{lesieur1996new, lesieur2005large} and Reynolds Stress Modeling (RSM) \cite{speziale1991analytical, launder1975progress, mishra2014realizability, speziale1991modelling}. The percentage of turbulence that is resolved and modeled at various degrees of accuracy varies across these methods \cite{speziale1991analytical, pope2001turbulent}. This also results in a scaling of the computational expenses, with LES being the most costly method in terms of computing. For a modest computing cost, Eddy Viscosity Models can forecast many complex flows with a reasonable degree of accuracy. Turbulence models based on eddy viscosity are the mainstay of flow simulations used in engineering design.

Eddy viscosity models maintain robustness and computational efficiency by utilizing significant simplifications in their formulation. Among them are the turbulent viscosity hypotheses (TVH) \cite{schmitt2007boussinesq} and the gradient diffusion hypothesis \cite{da2007analysis}. The amount that eddy viscosity models can capture the turbulent physics of intricate flows, including those with rotational effects, streamline curvature, flow separation, etc., is limited by these simplifications. Their predictions become questionable due to epistemic factors. 

The lack of correct physics incorporation in turbulence models and our poor understanding of turbulence physics lead to epistemic uncertainties in turbulent flow simulations \cite{smith2013uncertainty}. Turbulence models epistemic uncertainties can arise from a variety of sources, such as incomplete knowledge of the physics underlying turbulence \cite{duraisamy2019turbulence}, simplifications made to make the model computationally inexpensive \cite{alonso2017scalable}, lack of data to tune the model \cite{kato2013approach}, and simplifications made to make the model applicable for engineering workflows \cite{oliver2011bayesian, mishra2016sensitivity}. This epistemic uncertainty may be classified as parameter uncertainty owing to the estimated values of the closure coefficients in the model expressions and structural uncertainty related to the structure of the turbulence model expressions \cite{dow2011quantification}. When it comes to complicated real-life flows of engineering importance, turbulent flow simulations frequently have structural uncertainty as the primary source of mistakes and uncertainties. These flaws and uncertainties can have a considerable influence, resulting in considerably sub-optimal designs, in the iterative process of engineering design, when hundreds of intermediate designs are successively assessed using CFD simulations to optimize the final design. Therefore, trustworthy engineering designs require trustworthy structural uncertainty estimates from the turbulence model. The Design Under Uncertainty (DUU) method encompasses novel engineering design techniques such as resilient design and dependability based design \cite{yao2011review, padula2006aerospace}. The quality of the uncertainty estimates supplied to the technique will determine how well it performs. Therefore, accurate and calibrated estimates of the structural uncertainty of the turbulence model are crucial for engineering design.

The Eigenspace Perturbation Method (EPM) \cite{iaccarino2017eigenspace} is currently the sole data-free method for quantifying structural uncertainty in turbulence models. The eigenvalues, eigenvectors, and amplitude of the Reynolds stress tensor model are sequentially perturbed by the EPM. Discrete predictions are obtained by propagating CFD simulations via these altered Reynolds stresses. The structural uncertainty caused by the turbulence models is measured as the union over these forecasts. The EPM has been used in the recent past in a variety of engineering and science domains, such as civil engineering design \cite{gorle2019epistemic}, aerospace design and analysis \cite{mishra2019uncertainty, mishra2017rans, mishra2019estimating, mishra2017uncertainty, thompson2019eigenvector}, application to design under uncertainty (DUU) \cite{demir2023robust, cook2019optimization, mishra2020design, righi2023uncertainties}, virtual certification of aircraft designs \cite{mukhopadhaya2020multi, nigam2021toolset}, etc. 

Despite being widely used and having shown benefits, the EPM has drawbacks. The EPM's primary shortcoming is its dependence on physics alone. The EPM's ability to identify \textit{possible} states of turbulence in a given turbulent flow is limited by physics-based principles; it is not capable of identifying \textit{probable} states of turbulence for that particular turbulent flow. As an example, the EPM must equally weigh all states, including the limiting states, while analyzing the evolution of the anisotropy of the Reynolds stress tensor. For a homogenous flow where the 1- and 2-component limiting states are improbable, this is not ideal. Likewise, the EPM perturbs every point in the flow domain with the same value. The degree of perturbations is a measure of how much the high fidelity data and the predictions of the turbulence model disagree. Since the difference is not constant across the flow domain, the degree of perturbations also shouldn't be constant. These illustrations show how important it is to have a marker function that can distinguish between distinct turbulent flow instances and different sections of a turbulent flow domain in terms of the degree (or magnitude) of disturbances. 

With varying degrees of success, researchers have tried to create marker functions for the EPM. We speculate that since there is no analytical criterion to anticipate the difference of a RANS model prediction from the genuine development of the turbulent flow, it may not be possible to develop a marker function that is entirely based on physics. This marker function may be able to be approximated to arbitrary precision using a machine learning (ML) based method. In the past decade there have been a plethora of applications of machine learning in general and deep learning in particular across fields of science and engineering including material science \cite{morgan2020opportunities}, particle physics \cite{gupta2021improving}, energy sciences \cite{yao2023machine}, climate modeling \cite{rolnick2022tackling}, etc. Similarly numerous attempts to apply machine learning models and algorithms to problems related to fluid flows, turbulence modeling, combustion modeling, etc. have been successful recently \cite{duraisamy2019turbulence, brunton2020machine, chung2021data, chung2022interpretable, duraisamy2021perspectives, zhang2015machine}. Many of these approaches ignore the physics knowledge and instead concentrate on the ML model's approximation skills. Our aim is to enhance the EPM's foundation by utilizing machine learning models that are guided by domain expertise. Here, we use an affordable Convolutional Neural Network model for the marker function and highlight the lack of non-local modeling knowledge as a crucial delimiter. 

This manuscript is organized in a step-by-step fashion, with the first section outlining the main question that our approach attempts to answer. The EPM, the correction function, and the CNN model used in this study are all described in depth in the second part. Information on the test cases and data sets utilized is provided in the third section. The study's methodology is described in full in the fourth part. The findings and analysis of our investigation come next. An overview of the inquiry is provided at the end of the article.

\section{Mathematical Details}
\subsection{Eigenspace Perturbation Method (EPM)}
The idea of a turbulent viscosity is used by eddy viscosity models to close the evolution equation for the Reynolds stresses. The Boussinesq Hypothesis is another name for this \cite{pope2001turbulent}. It is assumed that the mean rate of strain tensor value and the instantaneous value of the Reynolds Stresses are directly related.

\begin{equation}
    \left\langle u_i u_j\right\rangle=\frac{2}{3} k \delta_{i j}-2 v_{\mathrm{t}}\left\langle S_{i j}\right\rangle,
\end{equation}

The turbulence kinetic energy is denoted by $k$, the Kronecker delta tensor by $\delta_{i j}$, the eddy viscosity coefficient by $\nu_{t}$, and the mean rate of strain by $\left\langle S_{i j}\right\rangle$. This supposition lowers the computing cost of engineering simulations by simplifying the RANS equations. However, for complicated flows, this assumption is quite restrictive and leads to erroneous eddy viscosity models. This linear connection, for instance, states that the Reynolds stress tensor and the mean rate of strain tensor have the same primary coordinates. For flows with turbulent separation or re-attachment, this is incorrect. Furthermore, this limits its applicability in situations where streamline curvature exists or when flows are dominated by rotation by ignoring any impact that the mean rate of rotation may have on the evolution of the Reynolds stresses.  

The EMP \cite{iaccarino2017eigenspace} inserts perturbations in the spectral representation of the Reynolds stress tensor predicted by the eddy viscosity model as a means of estimating the uncertainties resulting from simplifications in eddy viscosity-based modeling.

\begin{equation}\label{Eq:Rij_perturb}
        \left\langle u_{i} u_{j}\right\rangle^{*}=2 k^{*}\left(\frac{1}{3} \delta_{i j}+v_{i n}^{*} \hat{b}_{n l}^{*} v_{j l}^{*}\right).
\end{equation}

The perturbed eigenvalue matrix in this case is $\hat{b}_{k l}^{*}$, the perturbed eigenvector matrix is $v_{i j}^{*}$, and the perturbed turbulent kinetic energy is $k^{*}$. From a modeling point of view, the EPM substitutes the general relationship in which the eddy viscosity is a fourth-order tensor that takes into account the anisotropic character of turbulent flows for the linear, isotropic eddy viscosity assumption\cite{mishra2019theoretical}. 

\subsection{Correction function for RANS predictions}
Different methodologies are used in computational investigations of turbulent flows with respect to the specifics of the resolved and modeled scales. Every turbulent flow scale is resolved in DNS. DNS is not practical for engineering challenges and is very computationally costly. Smaller scales are modeled and certain scales are resolved in LES. Even if LES is less costly, the iterative design process employed in engineering design still requires too much computing power. Every turbulence scale is modeled in RANS. Thus, while RANS-based modeling is less costly than DNS or LES, it is not as high quality. We apply a correction term based on the difference between high fidelity DNS data and the predictions of the RANS model. This is how the correcting function works.

In this work, we propose to use a low-cost CNN-based model to improve the prediction accuracy of $k$, which is important for constructing $k^{*}$, as shown in Equation \ref{Eq:Rij_perturb}. We can express the outcomes of the RANS and DNS simulations as the following function of the perturbed turbulence kinetic energy $k^{*}$: 

\begin{equation}\label{Eq:Marker_Mk_Method}
    k^{*} = f(x,y).
\end{equation}

where $f$ is the function map from each coordinate $(x, y)$ to $k^{*}$, stated using a tuple $(x, y, k^{*})$ in simulation results. $x$ and $y$ are coordinates in a two-dimensional computational domain. The RANS correction function, in its simplest version, is a mapping between two functions:

\begin{equation}
Z: f^{\mathrm{RANS}}(x,y) \rightarrow f^{\mathrm{DNS}}(x,y)
\end{equation}

We can rewrite $Z$ as a mapping $\zeta$ between points that contains $k^{\mathrm{DNS}} = f^{\mathrm{DNS}}(x, y)$ and $k^{\mathrm{RANS}} = f^{\mathrm{RANS}}(x, y)$. The variables $f^{\mathrm{RANS}}$ and $f^{\mathrm{DNS}}$

\begin{equation}
\zeta: (x, y, k^{\mathrm{RANS}}) \rightarrow (x, y, k^{\mathrm{DNS}})
\end{equation}

With model error between RANS and DNS vis a vis kinetic energy, we get

\begin{equation}
   p^{\text {RANS }}\left(K_g \mid x, y\right)=p\left(k_g=k^{\text {RANS }} \mid x, y\right)
\end{equation}

\begin{equation}
    p^{\text {DNS }}\left(K_g \mid x, y\right)=p\left(k_g=k^{\text {DNS }} \mid x, y\right)
\end{equation}

where $K_g$ is the true kinetic energy at $(x, y)$. The DNS simulation produced kinetic energy. Kinetic energy from the RANS simulation $p^{\mathrm{DNS}}$ and its correction function $g$ may be used to estimate $p^{\mathrm{RANS}}$.

\begin{equation}
p^{\mathrm{DNS}}\left(K_g \mid x, y\right)=g\left(k^{\mathrm{RANS}}, x, y\right) p\left(k^{\mathrm{RANS}} \mid x, y\right)
\end{equation}

For every $x$, we have the following: $k_x^{\mathrm{DNS}} = f_x^{\mathrm{DNS}}(y)$ and $k^{\mathrm{RANS}} = f^{\mathrm{RANS}}(x, y)$, assuming both $f_x^{\mathrm{RANS}}$ and $f_x^{\mathrm{DNS}}$ are continuous. Stated differently, $(\mathbf{k}_{x,y,\delta}^{\mathrm{RANS}}, \mathbf{k}_{x,y,\delta}^{\mathrm{DNS}})$ pairs allow us to learn $\hat{g}$.

\subsection{CNN-based Correction Function}
In order to learn the correction function $\hat{g}$ from paired RANS and DNS simulation estimated kinetic energy $(\mathbf{k}_{x,y,\delta}^{\mathrm{RANS}}, \mathbf{k}_{x,y,\delta}^{\mathrm{DNS}})$, this inquiry uses a one-dimensional convolutional neural network (1D-CNN). Coordinates $(x, y)$ are only used to group neighbors of $k^{\mathrm{RANS}}$, so we grouped simulation data by $x$ and transformed $(y, k)$ at $x$ into $\mathbf{k}_{x,y,\delta}^{\mathrm{RANS}}$ via a rolling window parameterized by window size. This was necessary because our approximated correction function $\hat{g}$ depends on the neighbor of $k^{\mathrm{RANS}}$. With four layers and 86 parameters overall, our 1D-CNN is a single model for all zones at every $x$, correcting RANS toward DNS.

\section{Numerical Experiments}
This study's main goal is to modify the Eigenspace Perturbation Framework's application using convolutional neural networks that are computationally practical. Specifically, we aim to create a marker function that is capable of altering the eigenvalue perturbation throughout the computational domain and controlling its amplitude. In order to achieve this goal, we employ the CNN-based method to approximate this correction function for RANS simulations on two different datasets: the public RANS/DNS dataset \cite{voet2021hybrid} for two-dimensional channel flow over periodic hills, and the in-house RANS/DNS dataset \cite{zhang2021turbulent,chu2022model} for an SD7003 airfoil at $8^\circ$ angle of attack (referred to as \texttt{SD7003} in the following text).

\subsection{Data Flow and Model Configuration} \label{sec:VoetData}
\begin{figure}[h!]
    \centering
    \includegraphics[width=\linewidth]{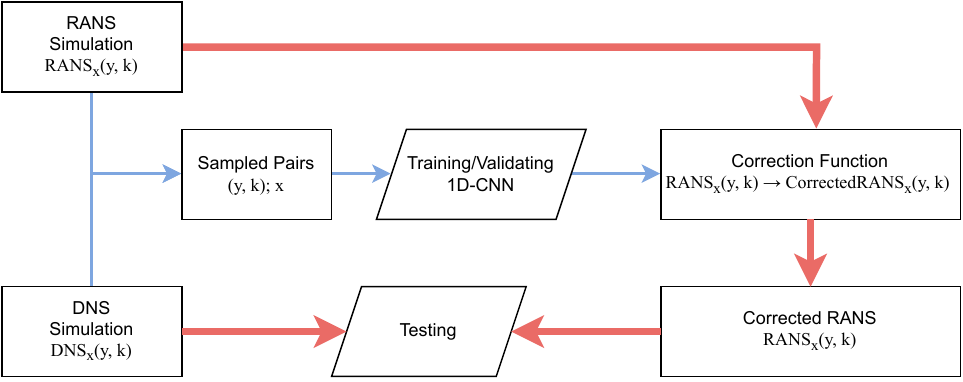}
    \caption{The flow of data \& logistical framework used in this study.} 
    \label{fig:data-flow.pdf}
\end{figure}

As shown in Figure \ref{fig:data-flow.pdf}, we configured trials of the 1D-CNN model for the two test scenarios after generating simulated data. The 1D-CNN corrected RANS prediction is assessed using the DNS data as the ground truth. The non-corrected RANS forecast serves as our reference point. Since the RANS and DNS frequently employ distinct grids, we choose $x$-coordinates that are shared by both types of meshes. The $x$-coordinate grouped pairs of $(\mathbf{k}_{x,y,\delta}^{\mathrm{RANS}}, \mathbf{k}_{x,y,\delta}^{\mathrm{DNS}})$ are divided into two sets: a validation set and a training set, according to the key $x$. For testing and training, we use an 80\%-20\% split of the raw dataset. 

The Mean Absolute Error (MAE) is the objective function that we utilize to train the model. Better final models were created in this work utilizing the MAE loss as the objective function since it does not punish inaccurate predictions as harshly as the Mean Squared Error (MSE or $L_2$ loss does. Both the uncorrected RANS and the 1D-CNN corrected RANS MAE loss are calculated. We compare these to demonstrate the effectiveness of our method.

\begin{figure*}[h!]
    \centering
    \includegraphics[width=\linewidth]{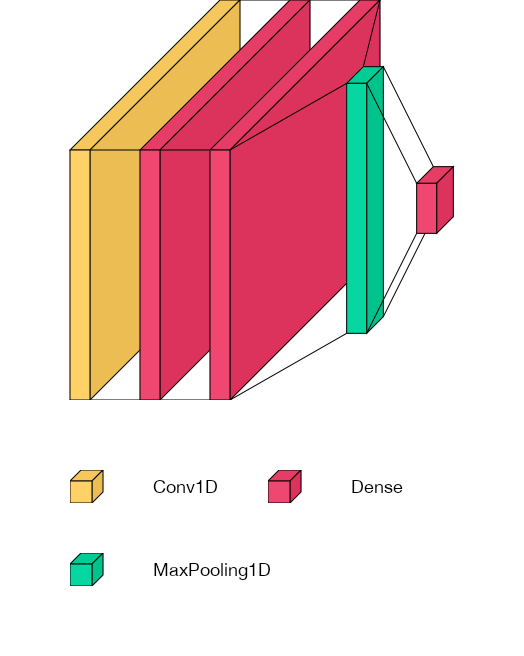}
    \caption{Details of the light-weight CNN utilized in this investigation} 
    \label{fig:model-layered-view.pdf}
\end{figure*}

The convolutional neural network architecture employed in this investigation is shown in Figure \ref{fig:model-layered-view.pdf}. For every DNS reading (window size of 1), we arrange the input from RANS with a window size of 11. The kernel size and stride of the 1-dimensional convolutional layer are set to 3. By using convolution procedures, this layer may extract representations of the high-level features from the incoming data. Two thick layers that interpret the feature that the earlier levels retrieved come after the convolutional layer. We have a Pooling layer (max pool) after them. The Adam optimizer with a learning rate of $0.001$ must be used for 800 epochs with a batch size of 10 in order to train the model.

\section{Results and Analysis }

We apply and validate the CNN technique performance at all paired (RANS, DNS) datasets at the common $x$ locations, as shown in the data flow diagram of the previous section. We give the findings in four sample $x$ axis spans in this section. The following locations span the separated region: $x/c = 0.17, 0.25, 0.32, 0.44$ for the \texttt{SD7003} dataset with RANS and DNS based on the airfoil geometry; and $x/H = 0, 0.035, 1.961, 4.885, 6.847$ for the \texttt{Voet} dataset based on the two-dimensional periodic hills.

\begin{figure}
    \centering
    \includegraphics[width=\linewidth]{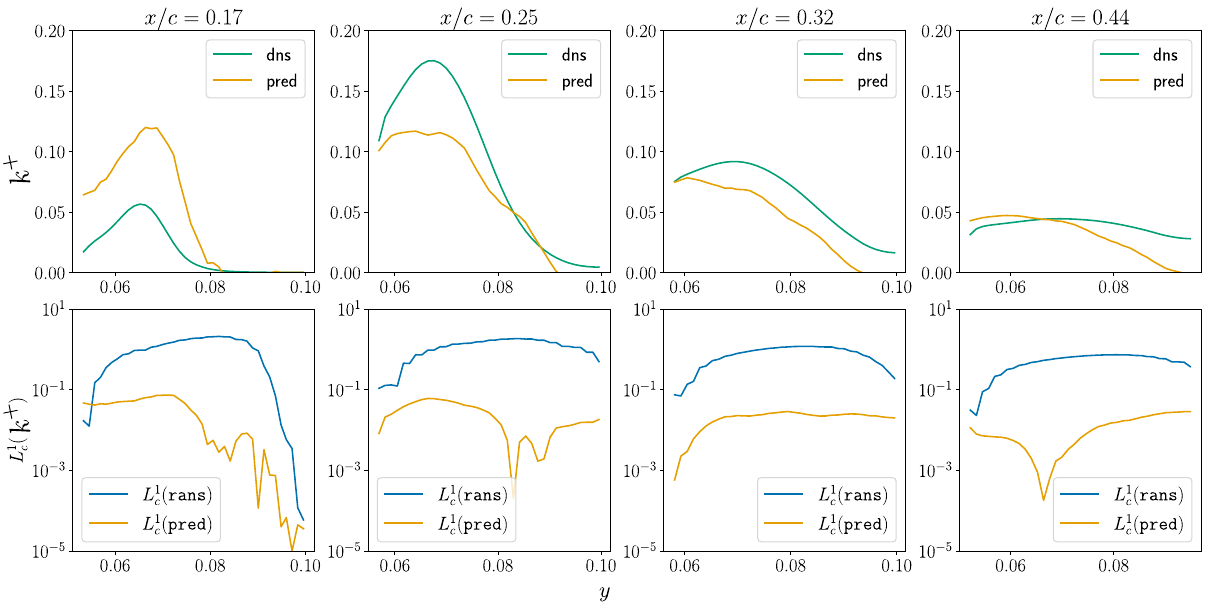}
    \caption{\texttt{SD7003} dataset results: CNN prediction for normalized tke. Top row: CNN predictions(\texttt{pred}) compared to true values (\texttt{dns}). Bottom row: CNN  error estimation by comparing MAE loss of $L^1_c(\texttt{rans})$ and $L^1_c(\texttt{pred})$.}
    \label{fig:rans-predicted-dns-main4-viz-loss.pdf}
\end{figure}

\begin{figure} 
    \centering
    \includegraphics[width=\linewidth]{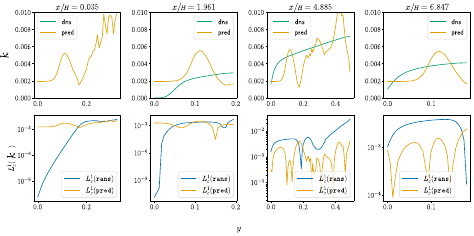}
    \caption{\texttt{Voet} periodic hills results: CNN prediction for dimensional tke. Top row: CNN Predictions compared to DNS case 7. For DNS: $\alpha = 0.8$ and $\gamma = 1.0$; DNS case 7: $\alpha = 1.2$ and $\gamma = 1.0$. Bottom row: Verror estimation by comparing MAE loss of $L^1_c(\texttt{rans})$ and $L^1_c(\texttt{pred})$. }
    \label{fig:case2-based-dns2}
\end{figure}

We present the findings for a test case of a flow over SD7003 airfoil and 2D periodic hills in the set of figures \ref{fig:rans-predicted-dns-main4-viz-loss.pdf} - \ref{fig:case2-based-dns2}. The moving average with a window size of six steps is used to smooth the CNN predicted profiles in the first row, as shown in Figures \ref{fig:rans-predicted-dns-main4-viz-loss.pdf} through \ref{fig:case2-based-dns2}. Despite the tiny amount of training data, the CNN prediction for the turbulent kinetic energy profile closely matches the ground truth DNS. For Figure \ref{fig:case2-based-dns2}, the turbulent kinetic energy is unscaled; for \ref{fig:rans-predicted-dns-main4-viz-loss.pdf}, on the other hand, the turbulence kinetic energy is normalized, that is, $k^{+} = k/U_{\infty}^2$. The CNN predicted DNS profiles are shown to resemble the qualitative aspects of the ground truth DNS profiles in Fig. \ref{fig:rans-predicted-dns-main4-viz-loss.pdf}.  We find a discrepancy in the segregated region at $x/c = 0.17$. The CNN-predicted DNS profiles get closer to the actual data as the flow moves downstream, indicating increased accuracy of the CNN correction function in the area. For every position, the corresponding $L^1_c(\texttt{pred})$ value exhibits a significant decline, roughly by two orders of magnitude, suggesting improved CNN prediction accuracy. An example of applying the CNN correction function to anticipate turbulent kinetic energy on the suction side of the SD7003 airfoil is shown in \ref{appendixA}. 

To train the CNN model for this inquiry, we performed further analysis on the publicly available RANS/DNS dataset \cite{voet2021hybrid}. We apply this model, trained on the RANS/DNS dataset case two \cite{voet2021hybrid}, to predict the $k$ profiles of DNS for case seven from the RANS/DNS dataset, as shown in Fig. \ref{fig:case2-based-dns2}. In instance 2, $\alpha = 0.8$ and $\gamma = 1.0$, but in case 7, $\alpha = 1.2$ and $\gamma = 1.0$, where $\gamma$ modifies the consecutive spacing of the hills and $\alpha$ modifies the steepness of the hills. For the sake of this comparison, turbulent kinetic energy is not scaled. The CNN predicted DNS profiles from Fig. \ref{fig:case2-based-dns2} resemble the ground truth DNS profiles accurately at all points, with the exception of $x/H = 0.034$, where a mismatch is seen. This is because of the extremely turbulent flow at this point, which is partly brought on by the large degree of separation. A significant divergence can be seen in the accompanying $L^1_c(\texttt{pred})$ along the wall at $x/H = 0.035$ and $x/H = 1.961$. This demonstrates the need for our CNN-based correction function to focus on improving prediction accuracy, especially in an area that is significantly divided. The total $L^1_c(\texttt{pred})$ value stays around 1-2 orders of magnitude lower than $L^1_c(\texttt{rans})$ as the flow moves downstream, indicating better forecasts. The incapacity of machine learning models to generalize from their training dataset to other, distinct flows is one of their main weaknesses. Our tests demonstrate the robustness and strong prediction retention of our CNN model across various training data sets.

In this study, we estimate the correction function that lessens the difference between RANS simulation and DNS data using a CNN model. This correction function provides direction on the amount of perturbation for $k^{*}$ necessary in Eqn. \ref{Eq:Rij_perturb}, reflecting the disparity in the RANS simulation results and acting as a spatially variable marker function for the eigenvalue perturbations. We test and implement this approach on two different datasets.

Despite the fact that the datasets belong to quite distinct flows, separation bubbles are present because of flow separation and re-attachment in both cases. The CNN model's corrective function may considerably lessen the difference between RANS predictions and DNS simulations in both of the flow test situations.

Our model was trained on a short dataset in both test conditions. When the CNN model was trained on one flow test case, it consistently produced accurate predictions on the second test flow instance, demonstrating its resilience to covariate alterations. Few research have been done to far on the use of ML models to create marker functions that help with the Eigenspace Perturbation Method by indicating the degree of difference in RANS predictions. Chu's recent work \textit{et al.} \cite{chu2022model} examined the application of polynomial regression to the kinetic energy of perturbed turbulence. The practical uses of this CNN model correcting technique include coupling it to the EPM. To provide a marker function for the perturbed turbulence kinetic energy, the EPM has been implemented inside the OpenFOAM framework \cite{chu2022model}. This CNN model correction technique will be applied as a spatially variable marker function to control the eigenvalue perturbation degree in further work.

\section{Summary}
This study aims to develop a spatially variable marker function that may adjust the EIgenspace Perturbation Framework's eigenvalue perturbation's size through the usage of deep convolutional neural networks. Using Ml models to modify and enhance the Eigenspace Perturbation Framework has been the subject of recent research \cite{heyse2021data, heyse2021estimating, matha2023evaluation}. That being said, this work is the first to use CNN models to assess the projection from the RANS prediction space to the DNS data space. We may incorporate non-local information into the uncertainty estimation through our usage of convolutions. The findings demonstrate that the CNN models develop a surrogate model for the marker function by learning the differences between RANS simulations and DNS data.

\appendix

\section{Application of the lightweight CNN-based correction function on UQ for an SD 7003 airfoil}\label{appendixA}
The study's Convolutional Neural Network model may be used to estimate the correction function for various flow scenarios after it has been trained. We describe its use in this instance for a particular flow situation.

\begin{figure*}[t]
         \centering
         \includegraphics[width=\textwidth, trim={2.4mm 0 4mm 0},clip]{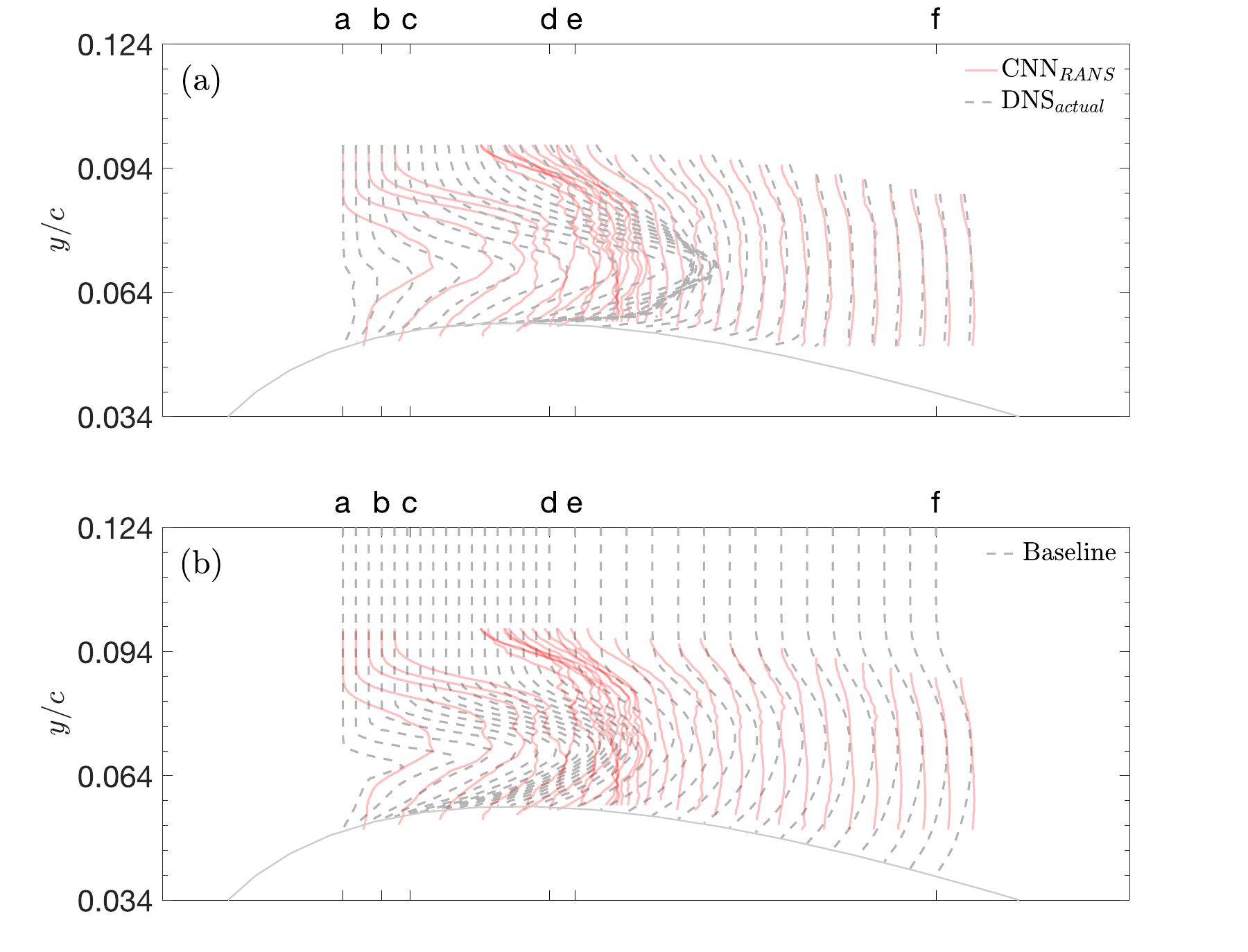}
        \caption{(a) Deep learning model corrected RANS (\texttt{CNN\_{RANS}}) {(solid-dotted lines)} of the normalized perturbed tke and (b) True data (\texttt{DNS\_{actual}}) on the suction surface of the SD7003 airfoil: left to right are zone $ab$, zone $cd$ \& zone $ef$. There are a total of thirty two points on the suction surface.}
        \label{fig:CNN_DNS.pdf}
\end{figure*}

In Figs. \ref{fig:CNN_DNS.pdf} (a) and (b), respectively, we provide the RANS profiles with the CNN model-based corrections in comparison to the baseline (uncorrected) RANS and the DNS (ground truth/high fidelity) profiles. For the $ab$ and $cd$ zones, the normalized turbulence kinetic energy profiles are evenly spaced with $x/c = 0.01$, whereas the $ef$ zone has a spacing of $x/c = 0.02$. The $ab$ and $cd$ zones have more densely packed normalized turbulence kinetic energy profiles because of the separation and re-attachment that occurs in this area. 
Based on Figures. \ref{fig:CNN_DNS.pdf} The CNN corrected RANS profiles in (a) and (b), which both display a progressive rise in the $ab$ and $cd$ zone, indicate a similar tendency to that for the ground truth dataset and the baseline RANS forecasts. Then, in the $ef$ zone farther downstream, a profile decrease is seen. 

Additionally, as the flow proceeds downstream, the size of the CNN-corrected RANS profiles generally increases in Fig. \ref{fig:CNN_DNS.pdf} (a), which is qualitatively comparable to the ground truth profiles. It is important to observe that near the beginning of the $ab$ zone, the CNN-corrected RANS profiles grow by a somewhat greater magnitude than the ground truth. As the flow continues downstream, the disparity steadily narrows, suggesting that further downstream our CNN model produces superior results. The $cd$ and $ef$ zones provide greater insight into this behavior. The ground truth profiles are grouped in the area where the $ef$ zone starts and the $cd$ zone ends because of the intricate flow characteristic of the reattachment \cite{chu2022quantification}. As seen in Fig. \ref{fig:CNN_DNS.pdf} (a), our CNN model effectively replicates this clustering tendency, albeit at a lower intensity than the ground truth. The normalized turbulence kinetic energy profiles in the $ef$ zone are often predicted by our CNN model to be accurate; that is, the CNN corrected RANS profiles and the ground truth profiles are almost identical. However, the comparison reveals an overall rather high disparity between the CNN-corrected RANS profiles and the baseline profiles across all zones in Fig. \ref{fig:CNN_DNS.pdf} (b). Furthermore, the ground truth DNS profiles reveal a noticeable clustering trend that is not present in the baseline profiles and is seen in the CNN-corrected RANS profiles. This implies that the existing CNN-based correction function performs better in precisely building $k^{*}$.

 \bibliographystyle{elsarticle-num} 
 \bibliography{cas-refs}





\end{document}